\documentstyle[12pt]{article}

\begin{document}
\thispagestyle{empty}
\vspace*{1cm}

\hfill{Preprint INR 924/96}

\hfill{June 1996}

\vspace*{2cm}

\begin{center}
{\bf MIXING OF NEUTRAL B MESONS AND FACTORIZATION\\}

\vspace*{2cm}
A.A.PIVOVAROV\\
{\it Institute for Nuclear Research, Russian Academy 
of Sciences\\
Moscow 117312, Russia\\}

\vspace*{2.5cm}

{\bf Abstract}  

\vspace*{0.5cm}

\parbox[c]{13cm}{A brief review of checking 
the factorization hypothesis for matrix element of $B^0 - \bar B^0$ mixing
within operator product expansion and QCD sum rules is given.  
Both perturbative and power corrections are considered.}

\vspace*{6cm}

\parbox[c]{13cm}{\it 
Talk given at III German-Russian Workshop on Theoretical Progress in 
}

{\it Heavy Quark Physics, Dubna, 20-22 May 1996}
\end{center}

\newpage
{\bf 1. Introduction}

Determination of the exact pattern of CP violation is one of the most
important problems of modern particle physics.
The system of neutral B mesons can provide some useful experimental
information on the subject. To decipher this information and convert
it into some knowledge of theoretical parameters of standard model or some
extended model one needs quite accurate theoretical calculation of
corresponding observables within the adopted theory.
At present an essential obstacle in getting precise theoretical estimates
for characteristics of CP violation related processes 
is a necessity of computing hadronic matrix elements that is
a completely nonperturbative problem. The most popular approximation 
for estimating such elements is
the factorization, or vacuum saturation, hypothesis the justification
of which is quite unclear. In the
present note we very briefly review some recent results of checking 
the validity of the factorization
hypothesis with analytical methods. We consider three point correlator
for computing power corrections violating the factorization approximation 
and two point correlator for
computing perturbative ones.

{\bf 2. Power corrections: three point correlator}

In order to develop a machinery of operator product expansion and QCD 
sum rules${}^{1,2,3}$ 
we use here a three point correlator of the form${}^{4,5}$
\begin{eqnarray}
\Pi_{\mu\nu}(p,p')&=&i^2\int dxdye^{ipx-ip'y}\langle 0|TJ_\mu(x)
O(0)_{\Delta B=2}J_\nu(y)|0\rangle \nonumber\\
&&=p_\mu p_\nu'\Pi_1(p^2,p'^2,q^2)+\ldots
=p_\mu q_\nu\Pi_2(p^2,p'^2,q^2)+\ldots
\label{bb31}
\end{eqnarray}
where
$q=p'-p$, $q^2=0$, $J_\mu=\bar d\gamma_\mu\gamma_5b$ 
is an interpolating current for B meson,
$$
\langle 0|J_\mu|B^0(p)\rangle=if_Bp_\mu.
$$
The operator $O_{\Delta B=2}$ is chosen with a standard normalization 
$$
O_{\Delta B=2}=
\bar b\gamma_\mu(1+\gamma_5)d\bar b\gamma_\mu(1+\gamma_5)d
$$
and the parameter $B_B$ is defined by the relation
$$
\langle \bar B^0|O_{\Delta B=2}|B^0\rangle={8\over 3}f_B^2 m_B^2 B_B.
$$
Within factorization approximation $B_B=1$.
For higher reliability we take for our analysis two invariant functions
$\Pi_{1,2}(p^2,p'^2,q^2)$ that appear in the expression for the three point 
correlator Eq.~(\ref{bb31}).
The dispersion relation 
dictates the following representation for these functions
after using saturation with the low lying resonance state  
\[
\Pi_i(p^2,p'^2)|_{q^2=0}=\int ds ds'{\rho(s,s')\over (s-p^2)(s'-p'^2)}
={8/3 f_B^4m_B^2 B_B\over (p^2-m_B^2)(p'^2-m_B^2)}+\ldots
\]

For both theoretical and physical parts of sum rules 
the factorization corresponds
to the following representation of the amplitudes
$$
\Pi_{\mu\nu}^{fact}(p,p')={8\over 3}T_{\mu\beta}(p)T_{\nu\beta}(p'),
$$
where 
$$
T_{\mu\beta}(p)
=i\int dxe^{ipx}\langle 0|TJ_\mu(x)\bar b(0)\gamma_\beta(1+\gamma_5)d(0)|0
\rangle,
$$
and gives the value $B_B=1$ 
for the parameter that describes the ratio of exact matrix 
element to the factorized one.

Leading contributions of power corrections 
that are not caught within the factorization 
approximation are given by some specific diagrams with external vacuum
fields. The first one is given by the gluon condensate${}^5$
$$
\Delta \Pi_1^G=
-{1\over 48\pi^2} \langle GG\rangle (pp')(5r(p^2)r(p'^2)+e(p^2) e(p'^2)),
$$
$$
\Delta \Pi_2^G=
\Delta \Pi_1^G
-{1\over 48\pi^2}\langle GG\rangle  p^2(r(p^2)r(p'^2)+e(p^2) e(p'^2))
+{1\over 24\pi^2}\langle GG\rangle  e(p^2) g(p'^2)
$$
where 
$$	
r(p^2)=\int_0^1dx {x\over -p^2x + m^2},
\quad
e(p^2)=\int_0^1dx {x-2x^2\over -p^2x + m^2},
$$
$$
g(p^2)=-\int_0^1 dx (1-2x) ln(1-xp^2/m^2),
$$
$m$ is the $b$ quark mass, 
$\langle GG\rangle = \langle {\alpha_s\over \pi}G_{\mu\nu}G^{\mu\nu}\rangle$.

Next contributions are due to condensates of operators with dimension
five in mass units. They are
$$
\Delta \Pi_1^\sigma=0,
\quad
\Delta \Pi_2^\sigma
=
-{m \langle g\bar d \sigma_{\mu\nu}G^{\mu\nu}d\rangle \over 6\pi^2 (p^2-m^2)}
\int_0^1 dx{x(x+1/2)\over -p'^2 x + m^2}.
$$
Finally, four quark operators contribute the amount that is not taken 
into account
by the factorization approximation for vacuum expectation values 
$$
\Delta \Pi_1^{4q}(p,p')=\Delta \Pi_2^{4q}(p,p')
=
-{2\langle
(\bar d\gamma_\mu(1+\gamma_5)d)^2\rangle^{non-fact}\over(p^2-m^2)(p'^2-m^2)}.
$$

As in most applications of operator product expansion we do not
include operators with dimension higher than six in our analysis.
Numerical results are
obtained after using Borel transformation in both $p^2$ and $p'^2$
independently and putting $M^2=M'^2$ afterwards. 
For both invariant functions there is a fairly wide window of
stability with respect to change of the Borel parameter. 
The numerical results are rather stable
and reveal only small violation of factorization. Namely, if  
$B_B=1+\Delta B_B$ then for both invariant functions and for wide
range of parameters ($m_0^2$, $\langle \bar q q \rangle$, \ldots) we
get${}^5$ 
$$
-\Delta B_B=0\div 0.1.
$$
This estimate is very conservative. The absolute value of
deflection from factorization approximation is about $-0.05$ for the
parameter $B_B$.
Main uncertainty is due to poor knowledge of numerical value of
$f_B$ on which there is a strong dependence (to the fourth power).

{\bf 3. Perturbative corrections: two point correlator}

Perturbative corrections of order $\alpha_s$ that violate the
factorization 
approximation are connected with genuine three loop massive diagrams.
Their computation with layout for sum rules technique based on three
point correlator with independent Borel transformation with regards
to kinematical variables $p^2$ and $p'^2$ can not be done at present
because of technical complexity. Therefore we turn to two point
correlator${}^{6,7}$ and introduce a quantity${}^{8}$
$$
T(x)
=\langle 0|TO_{\Delta B=2}(x)O_{\Delta B=2}(0)| 0 \rangle .
$$
The leading term of $\alpha_s$ expansion for 
the above correlator has an expression in
the configuration space that reads
$$
T_0(x)=2 N_c^2 \left(1+{1\over N_c}\right) 16
S'(x,m)S(-x,0)S'(x,m)S(-x,0)
$$
$$
=2 \left(1+{1\over N_c}\right)\mbox{tr}[S(x,m)S(-x,0)]
\mbox{tr}[S(x,m)S(-x,0)]
=2 \left(1+{1\over N_c}\right)\Pi_5(x)\Pi_5(x)
\eqno (2)
$$
where $S(x,m)$ is the free fermion propagator
and $N_c$ stands for the number of quark colors.
The prime means taking only the part of the propagator
that is proportional to a $\gamma$ matrix.
The function
$\Pi_5(x)=\langle 0|Tj_5(x)j_5(0)| 0 \rangle$
is the two point correlator
associated to the current $j_5 = \bar b i\gamma_5 d$.
Thus one observes a complete factorization in this order.

Eq.~(2) can be rewritten in the form
$$
T_0(x)
=2 \left( 1+{1\over N_c}\right) \Pi_{\mu\nu}(x)\Pi^{\mu\nu}(x),
\eqno (3)
$$
where
$\Pi^{\mu\nu}(x)=\langle 0|Tj_L^\mu(x)j_L^\nu(0)| 0 \rangle$
and
$j_L^\mu=\bar b_L \gamma^\mu d_L$.
The Lorentz decomposition in $x$-space reads
$$
\Pi^{\mu\nu}(x)
=(-\partial^\mu\partial^\nu+g^{\mu\nu}\partial^2)\Pi_T(x^2)
-\partial^\mu \partial^\nu\Pi_L(x^2)
$$
that again demonstrates an explicit factorization
in the configuration
space to leading order in $\alpha_s$.

The dispersion representation in $x$-space for any
two point correlator $\Pi_j(x)$ ($j=T,L,5$) has the form
$$
i\Pi_j(x^2)=\int_{s_j}^\infty r_j(s)D(x,s)ds
\eqno (4)
$$
where $D(x,s)$ is a free boson
propagator with the ``mass'' $\sqrt s$. 
The spectral functions $r_j$
read to leading order in $\alpha_s$
$$
r_L^{(0)}(s)={N_c\over 16 \pi^2}z(1-z)^2,
\quad r_T^{(0)}(s)={N_c\over 48 \pi^2}(1-z)^2(2+z),
\quad r_5^{(0)}(s)={m^2 N_c \over 8\pi^2} {(1-z)^2\over z}
$$
where $z=m^2/s$, $m$ is the $b$ quark mass.

The spectral function $\rho(s)$ of the
full correlator $T(x)$ is defined in the same way as in Eq.~(4).
To first order in $\alpha_s$ it can be expressed
in terms of the spectral functions $r_j(s)$ associated
to the two-line correlators in the form
$$
\rho(s)=\int r_1(s_1)r_2(s_2)\Phi(s;s_1,s_2)ds_1ds_2
$$
where
$$
\Phi(s;s_1,s_2)={1\over 16\pi^2 s}
\sqrt{(s-s_1-s_2)^2-4 s_1 s_2}
$$
is the two-body phase space factor.

To leading order in $1/N_c$, one can write the correlator $T(x)$
as a product of two two-line correlators. It is worthwhile to notice that
this decomposition is gauge invariant
and finite, i.e. it does not require any renormalization.

The full spectral density $\rho(s)$
has been computed numerically to the first
order in $\alpha_s$ ${}^8$ with a heavy use of known results for two
loop massive diagrams obtained earlier (e.g.${}^9$) and the
program of symbolic computation REDUCE. We analyze our results
concentrating on presentation of the
entire spectral density
$\rho(s)$ as a sum of factorizable and
nonfactorizable pieces
$$
\rho(s)
=
\rho_0(s)\left(1+\Delta \rho_f(s)+\Delta \rho_{nf}(s)\right).
$$
Nonfactorizable part of the spectral density is given by
one gluon exchange diagrams of a two-line correlator. 

Within sum rules approach one works with moments
$$
M_i(s_{th})=\int_{4m^2_b}^{s_{th}}\rho(s)s^{-i}ds
$$
that are decomposed as
$$
M_i(s_{th})=M_i^0(s_{th})
\left(
1+\Delta M_i^f(s_{th})+
\Delta M_i^{nf}(s_{th})
\right)
$$
according to the decomposition of
the spectral density.

A set of input parameters for numerical estimates is
$\Lambda^{(5)}_{\overline{MS}}=175~$MeV,
$m_b=4.6~$GeV.
The moments of the factorizable spectral density are
almost independent of the power $i$
of the weight function $s^{-i}$. The nonfactorizable
correction does not exceed a 15\% level with respect to the full
factorized spectral density.
Corrections of order $\alpha_s$ (factorizable+nonfactorizable)
are large for both the spectral density itself and its moments.
Depending on the energy $s$ they
can reach a magnitude of 100\% with respect to the leading
term. The nonfactorizable corrections
measured in terms of the fully factorized
(lowest order + $\alpha_s$ terms) spectral density
are moderate. 
All the factorizable corrections to the correlator, however,
can be absorbed into the calculation
of the decay constant $f_B$ from two point correlator with two
quark lines, in such a way that the relevant corrections to the $B_B$
parameter are only due to nonfactorizable ones.
Results for spectral density itself and for its moments 
for different values of integration regions are collected 
in Tables 1,2 taken from ref.${}^8$.

\begin{table}
\begin{center}
\begin{tabular}{|c|c|c|c|} \hline

$s/m_b^2$   &$\Delta\rho_f$ &$\Delta\rho_{nf}$
&$\Delta\rho_{nf}/(1+\Delta\rho_f)$\\ \hline

5.5 &1.03   &0.02     &0.01 \\ \hline
6.0 &0.95   &0.21     &0.11  \\ \hline
6.5 &0.89   &0.29     &0.15  \\ \hline
7.0 &0.84   &0.32     &0.17  \\ \hline
\end{tabular}
\caption{Normalized spectral densities}
\vspace{1cm}
\begin{tabular}{|c|c|c|c|c|} \hline

$i$&$s_{th}/m_b^2$&$\Delta M_f$ &$\Delta M_{nf}$ &$\Delta
M_{nf}/(1+\Delta M_f)$\\ \hline

0&5.5   &1.07 &-0.16 &-0.08  \\ \cline{2-5}
 &6.0   &0.99  &0.11  &0.06  \\ \cline{2-5}
 &6.5   &0.93  &0.23  &0.12  \\ \cline{2-5}
 &7.0   &0.88  &0.29  &0.15  \\ \hline

5&5.5   &1.08 &-0.21 &-0.10  \\ \cline{2-5}
 &6.0   &1.00  &0.08  &0.04  \\ \cline{2-5}
 &6.5   &0.94  &0.20  &0.11  \\ \cline{2-5}
 &7.0   &0.89  &0.27  &0.14  \\ \hline

10&5.5   &1.09 &-0.27 &-0.13  \\ \cline{2-5}
  &6.0   &1.01  &0.02  &0.01  \\ \cline{2-5}
  &6.5   &0.96  &0.16  &0.08  \\ \cline{2-5}
  &7.0   &0.91  &0.23  &0.12  \\ \hline
\end{tabular}
\end{center}
\caption{Normalized moments of spectral densities}
\end{table}

\end{document}